\documentclass[pdflatex,sn-mathphys-num]{sn-jnl}% Math and Physical Sciences Numbered Reference Style 
\usepackage{graphicx}%
\usepackage{multirow}%
\usepackage{amsmath,amssymb,amsfonts}%
\usepackage{amsthm}%
\usepackage{mathrsfs}%
\usepackage[title]{appendix}%
\usepackage{xcolor}%
\usepackage{textcomp}%
\usepackage{manyfoot}%
\usepackage{booktabs}%
\usepackage{algorithm}%
\usepackage{algorithmicx}%
\usepackage{algpseudocode}%
\usepackage{listings}%
\usepackage{url}
\usepackage{lipsum}
\usepackage{hyperref}
\theoremstyle{thmstyleone}%
\theoremstyle{thmstyletwo}%
\theoremstyle{thmstylethree}%

\begin{document}

\title[Article Title]{A Comprehensive Framework for Building Highly Secure, Network-Connected Devices: Chip to App}


\author[]{\fnm{Khan} \sur{Reaz}}
% \email{khanreaz@ieee.org}

\author[]{\fnm{Gerhard} \sur{Wunder}}
% \email{g.wunder@fu-berlin.de}


\affil[]{\orgdiv{Dept. of Mathematics and Computer Science},\orgname{ Freie Universität Berlin}, \orgaddress{ \country{Germany}}}

% \affil[2]{\orgdiv{Department}, \orgname{Organization}, \orgaddress{\street{Street}, \city{City}, \postcode{10587}, \state{State}, \country{Country}}}

% \affil[3]{\orgdiv{Department}, \orgname{Organization}, \orgaddress{\street{Street}, \city{City}, \postcode{610101}, \state{State}, \country{Country}}}


\abstract{
The rapid expansion of connected devices has amplified the need for robust and scalable security frameworks. This paper proposes a holistic approach to securing network-connected devices, covering essential layers: hardware, firmware, communication, and application. At the hardware level, we focus on secure key management, reliable random number generation, and protecting critical assets. Firmware security is addressed through mechanisms like cryptographic integrity validation and secure boot processes. For secure communication, we emphasize TLS 1.3 and optimized cipher suites tailored for both standard and resource-constrained devices. To overcome the challenges of IoT, compact digital certificates, such as CBOR, are recommended to reduce overhead and enhance performance. Additionally, the paper explores forward-looking solutions, including post-quantum cryptography, to future-proof systems against emerging threats. This framework provides actionable guidelines for manufacturers and system administrators to build secure devices that maintain confidentiality, integrity, and availability throughout their lifecycle.
}




\maketitle

\section{Introduction}
The growing prevalence of network-connected devices in diverse industries such as healthcare, manufacturing, and smart infrastructure has significantly increased security concerns. Devices forming the Internet of Things (IoT) ecosystem are often resource-constrained, making them vulnerable to a wide range of attacks, including man-in-the-middle (MitM) attacks, ransomware, and side-channel exploits. To ensure the safety and reliability of these systems, a holistic approach to network security is essential.

This paper introduces a comprehensive security framework that addresses four critical layers of network-connected devices:

\begin{itemize}
    \item[] \textbf{Hardware}: Ensuring secure key management, random number generation, and hardware-based root-of-trust mechanisms.

    \item[] \textbf{Firmware}: Enforcing cryptographic integrity validation, secure boot processes, and protection against tampering.

    \item[] \textbf{Communication}: Implementing Transport Layer Security  and compact certificates optimized for resource-constrained environments.

    \item[] \textbf{Application}: Achieving end-to-end encryption and message integrity validation while maintaining performance efficiency.
\end{itemize}

In addition to current security standards, we explore emerging post-quantum cryptographic solutions to future-proof connected devices against potential quantum computing attacks. These recommendations are designed to provide a practical, scalable, and forward-looking approach to securing the next generation of network-connected systems.
The remainder of the paper is organized as follows: Section 2 defines key terminologies and requirements for secure network design.  Section 3 outlines essential cryptographic concepts and attack scenarios. Section 4 provides actionable recommendations for secure implementation across hardware, firmware, and network layers. Section 5 compares current cryptographic standards with post-quantum cryptographic advancements. Finally, Section 6 concludes the paper and highlights future research directions.

\section{Terminologies and Requirements}\label{term}
This section introduces the foundational terms and requirements essential for designing and implementing secure network-connected devices.

\subsection{Components}
\label{sec:2:components}

\subsection*{User}
The user, or often known as end-user, is a human entity who is legally registered to some authority by proving their identity. 

\subsection*{Device}
A device refers to any physical entity equipped with computational capabilities that can perform data processing, communication, and interaction within a network. Devices encompass a broad range of hardware, from traditional computing systems like desktops and servers to mobile phones, IoT sensors, and embedded systems.

\subsection*{Authenticator}
An authenticator is a handheld smart device, such as a smartphone or tablet, equipped with a rich user interface and capable of communication through multiple radio interfaces, including Wi-Fi, cellular networks, and Bluetooth.

\subsection*{Digital ID/Smart eID}
A digital ID, or smart eID, functions as the digital counterpart of a physical ID card, stored on a smart device.

\subsection*{eID Service}
The eID service fulfills several critical functions, including managing communication with the ID card's chip, retrieving and updating authorization certificates and revocation lists, and securely transmitting verified data.

\subsection*{eID Service Provider}
Service providers, such as hospitals, hotels, and public transportation companies, can operate their own eID servers by obtaining accreditation from the eID issuer.

\subsection*{eID Server}

The eID server comprises both hardware and software components operated by the service provider to integrate eID functionality into its IT infrastructure. Its primary functions include establishing secure communication with client software and the ID card’s chip, transmitting data to the specified service, verifying the ID card’s validity and authenticity, checking for any blocks imposed by the ID card holder, and communicating the eID function results to the provider’s other systems.\\

Additionally, the eID server regularly receives updated authorization certificates and revocation lists from the certification provider, ensuring ongoing compliance and security. Service providers wishing to develop their own eID servers must ensure that both hardware and software components comply with the Technical Guidelines established by the Federal Office for Information Security. This compliance is crucial for executing cryptographic protocols with the ID card’s chip and for routinely receiving necessary authorization certificates and revocation lists.

\subsection*{Chip Authentication}

The Chip Authentication process establishes a secure connection between RF chips and readers and enables detection of cloned RF chips within identification documents. Each RF chip supporting this method possesses a unique public-private key pair, with the private key securely stored in a protected area, inaccessible even in the event of cloning.\\

During the Chip Authentication process, the RF chip transmits its public key along with a random number to the reader, which subsequently generates its own key pair. By using their respective private keys, the exchanged public key, and the random number, both the RF chip and reader compute a shared secret key. This shared key provides robust encryption for all subsequent communication, thereby ensuring data security. The reader then authenticates the chip’s private key through this shared secret. If a cloned RF chip attempts to generate a new set of keys, Passive Authentication can detect this alteration, as the public key is protected from unauthorized modification by a digital signature.

\subsection*{Passive Authentication}

Passive Authentication (PA) is a process used to verify the authenticity and integrity of the data stored on the eID chip.\\

When an eID is issued, its data is digitally signed using a Document Signer certificate, provided by the Country Signing Certificate Authority (CSCA) of the issuing country. Access to this certificate is restricted to entities authorized by the issuing nation, ensuring that only validated data is signed and stored on the eID. The CSCA certificate acts as the root of the country’s public key infrastructure (PKI), establishing a certificate hierarchy that certifies the legitimacy and accuracy of data on official identification documents.\\

During the eID verification process, Passive Authentication confirms the digital signature on the chip’s data, linking it back to the CSCA certificate. This procedure verifies that the information on the RF chip is both authentic and securely issued by an authorized document producer.

\subsection*{Server/Back-end}

An IoT analytics provider may operate its services within its own cloud infrastructure or within a hosted cloud environment. The physical or virtual instance of this service is referred to as the server $(S)$. This server is accessible via the internet, or, in the case of a local installation, it can be discovered within the local network. Access to the server's APIs is secured through authentication and authorization mechanisms, ensuring secure communication and reliable data handling.

\subsection*{Enrollee}

An Enrollee is a newly manufactured device that has not yet been provisioned. It is equipped with IP-based network connectivity and includes at least one wireless interface, such as Wi-Fi, Ultra-Wideband, or Ethernet. This interface supports secure communication during the initial setup and onboarding process.

\subsection*{Gateway/Router}

A Gateway or Router manages data traffic within the local network and provides a connection bridge to external networks. Accessible through IP-based connections, it serves as both a routing and network access point.\\

Beyond routing data, the gateway enforces security measures such as authentication, firewalling, and access control, ensuring that network traffic is both efficiently managed and secure.

\subsection*{System Administrator}
A System Administrator (SysAdmin) is responsible for managing and maintaining an organization’s core IT infrastructure, ensuring the security, reliability, and optimal performance of servers, networks, and connected devices. The SysAdmin oversees tasks such as configuring and deploying servers, managing user access, maintaining software updates, and implementing security protocols.\\

In the context of cryptographic operations, the SysAdmin plays a vital role in initiating and managing key processes, including generating cryptographic key pairs, creating and submitting Certificate Signing Requests (CSRs), and installing certificates on the appropriate servers.

\section{Definitions}
\label{sec:2:definition}
This section introduces fundamental cryptographic concepts and highlights critical attack types that secure network architectures must withstand.\\

% \subsection*{Identity }

\noindent \textbf{Universally unique identifier (UUID)}  is a 128-bit number used to identify information in computer systems. The term globally unique identifier (GUID) is also used.\\

% \subsection*{Cryptography}

\noindent A \textbf{root of trust (RoT)} is a set of functions that are always trusted by an operating system. It serves as the foundation for all secure operations in a computing system. A RoT contains keys used for digital signing and verification, as well as cryptographic functions that enable secure boot processes. It is an essential security asset. Embedding an RoT in hardware provides a trusted execution environment and creates a solid foundation for electronic systems security.\\

\noindent \textbf{Key} is a piece of information (for example, a randomized bit-string) required to encrypt or decrypt data.\\

\noindent \textbf{Key management system (KMS)} is used to manage cryptographic keys, including their generation, storage, use, rotation, destruction, and replacement.\\

\noindent \textbf{Key type} is a specific implementation type for a cryptographic primitive.\\

\noindent \textbf{Plaintext}: Any infromation in its original form.\\

\noindent \textbf{Primitive} is  cryptographic building block that manages an underlying algorithm so users can perform cryptographic tasks safely. \\

\noindent \textbf{Symmetric key encryption}: A cryptographic algorithm that uses the same key to encrypt plaintext and decrypt ciphertext.\\

\noindent \textbf{Asymmetric key encryption}: A cryptographic system that uses paired keys—public and private—to encrypt and decrypt data. Public keys are used to encrypt data and may be shared. Private keys are used to decrypt data, and are only known to the owner.\\

\noindent \textbf{Ciphertext}: The result of encryption performed on plaintext using an algorithm. Ciphertext is not understandable until it has been converted back into plaintext using a key.\\

\noindent \textbf{Deterministic encryption}: A type of encryption that always produces the same ciphertext for a given plaintext and key. This can be risky, because an attacker only needs to find out which ciphertext corresponds to a given plaintext input to identify it.\\

\noindent \textbf{Hybrid encryption}: A cryptographic system that combines asymmetric key encryption and symmetric key encryption. Hybrid encryption combines the efficiency of symmetric encryption with the convenience of public-key encryption. To encrypt a message, a fresh symmetric key is generated and used to encrypt the plaintext data, while the recipient’s public key is used to encrypt the symmetric key only. The final ciphertext consists of the symmetric ciphertext and the encrypted symmetric key.\\

\noindent \textbf{Adaptive Chosen Ciphertext Attack}: An adversary with access to a decryption functions attempts to defeat the security of a scheme to which the function belongs. The attacker has the capability to continually access the oracle and get a response for polynomially many arbitrary ciphertext.\\

\noindent \textbf{Perfect forward secrecy } is commonly used to denote a feature of key agreement protocols which gives assurances that past session keys will not be compromised even if the private key  is compromised.\\

\noindent \textbf{Digital Signature} confirms the authenticity and the integrity of the data by signing it with key. It uses asymmetric keys: private key is used for signing and public key is used for verifying.\\

\noindent \textbf{Zero-Trust Model} or Zero-Trust architecture is a collection of concepts designed to minimise uncertainty in enforcing accurate access decisions in compromised networks. \\

\noindent \textbf{Phishing attacks}: These attacks involve an attacker sending a fake email or message that appears to be from a legitimate source, such as a bank or social media platform, with the intention of tricking the recipient into sharing sensitive information, such as passwords or financial details.\\

\noindent  \textbf{Malware attacks}: Malware is a type of software designed to damage or gain unauthorized access to a computer system. Examples of malware include viruses, Trojan horses, and ransomware.\\

\noindent  \textbf{Denial of Service (DoS) attacks}: These attacks involve flooding a network or website with traffic, making it unavailable to users. Distributed Denial of Service (DDoS) attacks involve multiple sources flooding the targeted network or website.\\

\noindent  \textbf{Man-in-the-middle (MitM) attacks}: These attacks involve an attacker intercepting communications between two parties and potentially altering the information being exchanged.\\

\noindent  \textbf{Password attacks}: These attacks involve attempting to guess or crack passwords to gain unauthorized access to a system or account.\\

\noindent  \textbf{SQL injection attacks}: These attacks involve exploiting vulnerabilities in a website or application's code to gain unauthorized access to its underlying database.\\

\noindent \textbf{Cross-site scripting (XSS) attacks}:  insert malicious code into a legitimate website or application script to get a user's information, often using third-party web resources. Attackers frequently use JavaScript for XSS attacks, but Microsoft VCScript, ActiveX and Adobe Flash can be used, too.\\

\noindent  \textbf{DNS tunneling} is used by cybercriminals to exchange application data, like extract data silently or establish a communication channel with an unknown server.\\

\noindent \textbf{Backdoor Trojan} creates a backdoor vulnerability in the victim's system, allowing the attacker to gain remote, and almost total, control. Frequently used to link up a group of victims' computers into a botnet or zombie network, attackers can use the Trojan for other cybercrimes.\\

\noindent \textbf{Ransomware} is sophisticated malware that takes advantage of system weaknesses, using strong encryption to hold data or system functionality hostage. Cybercriminals use ransomware to demand payment in exchange for releasing the system. A recent development with ransomware is the add-on of extortion tactic.\\

\noindent \textbf{Zero-day exploit} attacks take advantage of unknown hardware and software weaknesses. These vulnerabilities can exist for days, months or years before developers learn about the flaws.\\

\noindent \textbf{p-value} is a statistical measure that quantifies the probability of observing a test value that is at least as extreme as the particular value that has just been observed (tail probability) if the null hypothesis is true. If this p-value is smaller than a pre-defined bound, it indicates that the null hypothesis should be rejected. \\

\noindent A \textbf{botnet}  is defined as a group of internet-connected devices that have been subjected to a hacking attack and are subsequently controlled as a collective without the knowledge or consent of the device owner. These devices are typically used for the purpose of carrying out malicious activities. \\

\noindent \textbf{Confidentiality}  ensures that information is only accessible to authorized parties and is kept secret from unauthorized entities.

\noindent {Example:} Encryption ensures that only intended recipient(s) can read the content.\\

\noindent \textbf{Integrity} guarantees that the data has not been altered or tampered with during transmission or storage.

\noindent {Example:} Message Authentication Codes (MACs), digital signatures, and hash functions are commonly used to verify integrity.\\

\noindent \textbf{Authentication}  checks that the parties involved in communication are who they claim to be.

\noindent {Examples:} Passwords, biometrics, and digital certificates\\

\noindent \textbf{Authorization} determines whether an authenticated entity has permission to perform a specific action or access specific resources.

\noindent {Examples:} Role-based access control (RBAC) and Access control lists (ACLs).\\

\noindent \textbf{Non-repudiation} makes sure that a party cannot deny the authenticity of their signature on a document or a message that they originated.

\noindent {Examples:} Digital signatures, logging mechanisms.\\

\noindent  \textbf{Availability} means that systems and data are available to authorized users when needed, preventing disruptions from attacks or failures.

\noindent {Examples:} Redundancy and failover systems, protection against Denial of Service (DoS) attacks.\\

\noindent \textbf{Forward Secrecy} dictates that session keys will not be compromised even if long-term keys are compromised in the future.

\noindent {Examples:} Advanced Diffie-Hellman key exchange in protocols like TLS.\\

\noindent\textbf{Anonymity}  means that the identity of a user remains hidden from other users or third parties.

\noindent {Examples:} Anonymity networks like Tor, anonymous credentials.\\

\noindent\textbf{Unlinkability} occurs when an adversary cannot link two or more actions, messages, or identities to a single entity.

\noindent {Examples:} Mix networks, onion routing.\\

\noindent\textbf{Unobservability} ensures that an adversary cannot detect whether a particular action or communication is taking place.

\noindent {Examples:} Masking channels.\\

\noindent \textbf{Accountability} ensures that actions of individuals or systems can be traced to them, and they can be held responsible for their actions.

\noindent {Examples:} Logging and auditing systems, cryptographic signatures.\\

\noindent \textbf{Verifiability}  ensures that the correctness of operations or transactions can be independently verified by any party.

\noindent {Examples:} Verifiable voting systems, publicly verifiable encryption schemes.\\

\noindent \textbf{Auditability}   ensures that actions and events can be recorded and reviewed, typically for security, compliance, or forensic purposes.

\noindent {Examples:} System logs, blockchain technology.\\

\noindent \textbf{Trustworthiness} ensures that the system or component can be relied upon to perform its intended function securely and correctly.

\noindent {Examples:} Trusted platform module (TPM), formal verification of protocols.\\

\noindent \textbf{Robustness}  ensures that the system can withstand attacks, failures, or environmental changes without compromising security.

\noindent {Examples:} Fault-tolerant systems, resilience against side-channel attacks.\\

\noindent \textbf{Resistance to Side-Channel Attacks} means that the system is resistant to attacks that exploit unintended information leakage, such as timing, power consumption, or electromagnetic emanations.

\noindent {Examples:} Constant-time algorithms, shielding and noise injection.\\

\noindent \textbf{Resistance to Replay Attacks} means that an attacker cannot reuse a previously intercepted message to deceive the system or its participants.

\noindent {Examples:} Nonces, timestamps, sequence numbers.\\

\noindent \textbf{Resistance to Impersonation Attacks} ensures that an attacker cannot successfully masquerade as another user or entity.

\noindent {Examples:} Multi-factor authentication, Public key infrastructure (PKI).\\

\noindent \textbf{Resistance to Collusion} ensures that even if multiple adversaries or compromised entities collude, they cannot break the security of the protocol.

  \noindent {Examples:} Secret sharing schemes, distributed consensus protocols.\\

\noindent \textbf{Resilience to Denial-of-Service (DoS) Attacks} ensures that the system remains operational or degrades gracefully under attack conditions.

\noindent {Examples:} Rate limiting, CAPTCHA, traffic analysis.\\

\noindent \textbf{Ephemeral Key Security} ensures that the security of a session is not compromised if the ephemeral (temporary) keys are exposed or lost.\\

\noindent \textbf{Resistance to Cryptanalysis} ensures that cryptographic schemes are resistant to attacks that aim to break the encryption, such as brute-force attacks, differential cryptanalysis, or algebraic attacks.

\noindent {Examples:} Use of strong cryptographic algorithms  ChaCha20, continual updates to cryptographic standards.\\

 \noindent \textbf{Resistance to Traffic Analysis} means that an adversary cannot gain useful information by analyzing patterns in the flow of communications.

\noindent {Examples:} Padding, randomization techniques, onion routing.\\

\noindent \textbf{Formal Verification and Proof of Security} ensures that the security properties of a protocol or system can be mathematically proven, usually within a formal model or framework.

\noindent {Examples:} Formal methods, proof assistants like ProVerif or Tamarin.\\

\noindent \textbf{Indistinguishability under Chosen Plaintext Attack (IND-CPA)} formalizes the security of an encryption scheme against an adversary who can choose plaintexts to be encrypted. The scheme is considered secure if the adversary cannot distinguish between the encryptions of two chosen plaintexts, even when they have access to an encryption oracle that provides encryptions of plaintexts of their choice. Essentially, even if the adversary can see the ciphertexts of any plaintexts they select, they should not be able to gain any advantage in distinguishing between the encryptions of two specific plaintexts.\\
    
\noindent \textbf{Indistinguishability under Chosen Ciphertext Attack (IND-CCA)} formalizes the security of an encryption scheme against an adversary who can adaptively choose ciphertexts to be decrypted, except for the challenge ciphertext. The scheme is considered secure if the adversary cannot distinguish between the encryptions of two chosen plaintexts, even with access to a decryption oracle.\\

\noindent \textbf{Indistinguishability under Adaptive Chosen Ciphertext Attack (IND-CCA2)}  provides a formal framework for ensuring the security of an encryption scheme against an adversary who can adaptively choose ciphertexts to be decrypted, both before and after receiving a challenge ciphertext, except for the challenge ciphertext itself. The scheme is deemed secure if the adversary is unable to distinguish between the encryptions of two selected plaintexts, even with adaptive access to both encryption and decryption oracles. This implies that, despite having the capability to decrypt any ciphertexts (except the challenge), the adversary cannot confirm any information that would assist them in differentiating between the encryptions of the two chosen plaintexts.

\section{Recommendations on procedures}
\label{sec:2:requirements}

In addition to presenting security protocols, this work aims to offer a set of clear guidelines and technical requirements to establish a balanced level of cryptographic security applicable to all network-connected devices manufactured by device producers, regardless of industry. We have identified \emph{\textbf{five}} essential characteristics for achieving highly secure, network-connected devices:\\

\begin{figure}[htbp]
    \centering
    \includegraphics[width=0.5\linewidth]{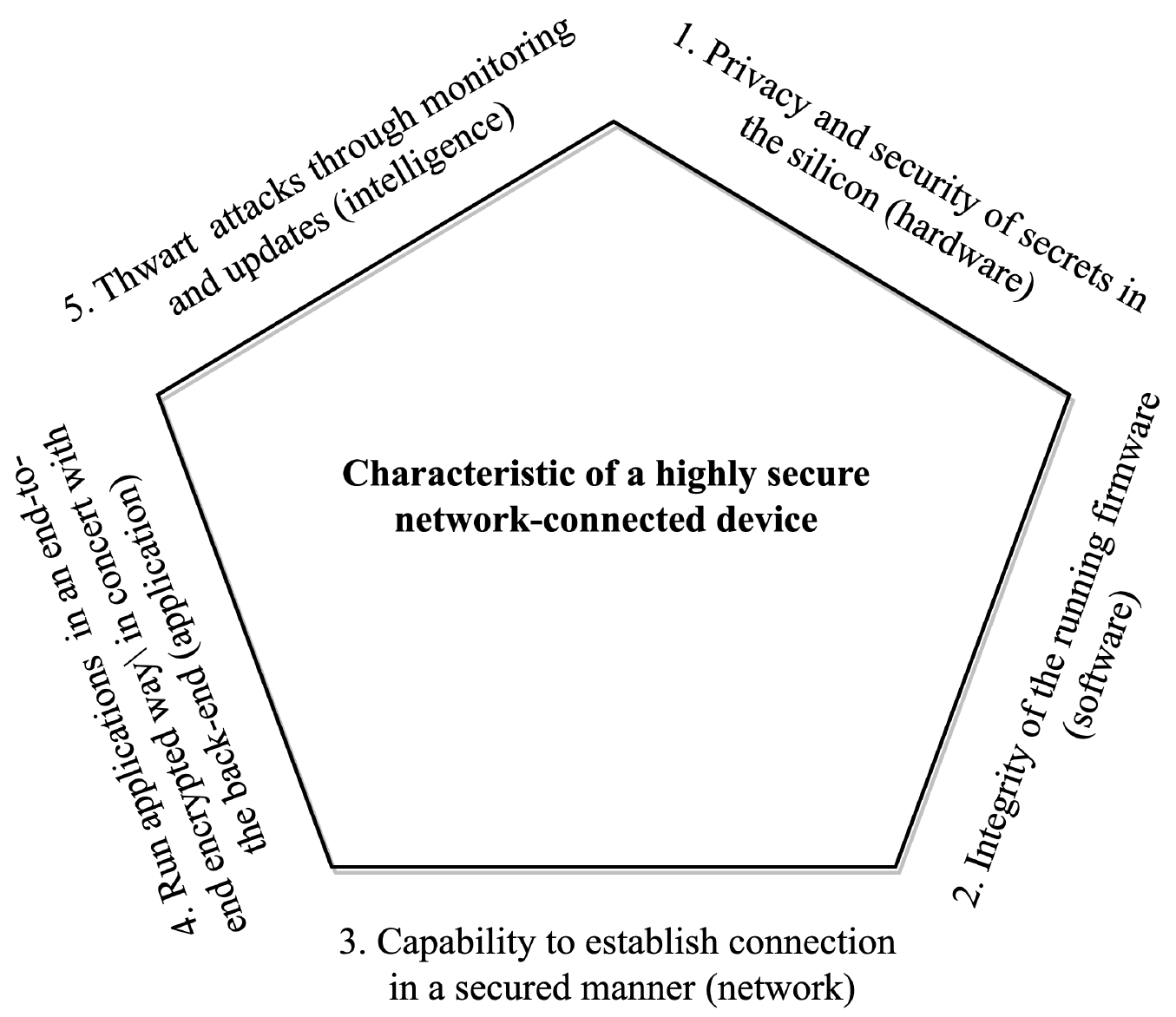}
    \caption{Five characteristics of a secure-networked device}
    \label{fig:enter-label}
\end{figure}

\begin{enumerate}

    \item \emph{Privacy and security of secrets in the  silicon \textbf{(hardware)}}    
    \item \emph{Integrity of the running firmware \textbf{(software)}}
    \item \emph{Capability to establish connection in a secured manner \textbf{(network)}}
    
    \item \emph{Run \textbf{applications}  in an end-to-end encrypted way }

    \item \emph{Thwart  attacks through \textbf{intelligent} monitoring and updates }

\end{enumerate}
In this work, we concentrate on the first four key areas: hardware, firmware, network, and applications. The fifth area, intelligence, is reserved for future research.

\subsection{  Random Number Generation}

In industry, two primary standards guide the design and evaluation of both deterministic and non-deterministic random number generators (RNGs): the guidelines provided by the National Institute of Standards and Technology (NIST)~\cite{nist80090a} and the AIS 20/31 specification issued by the German Federal Office for Information Security (BSI)\cite{BSI_RNG_2024}. It is recommended that readers consult these specifications to verify compliance of the RNG implementation on a given device.

\subsection{Cryptographic assets generation and storage }

The generation and injection of a device root key into a SoC typically occur during the manufacturing process. This procedure must include the following essential steps:

\begin{enumerate}

\item[--] A RNG creates a unique private key for each device. The RNG relies on a high-entropy source, such as a hardware-based true random number generator (TRNG), to ensure that the key is random and unique.

\item[--] A cryptographic algorithm, such as ECC, is used to generate a corresponding public key from the private key.

\item[--] The public key is then signed by a trusted authority, such as a CA, resulting in a digital certificate that verifies the authenticity of the device.

\item[--] The signed digital certificate is injected into the device during manufacturing and is typically stored in non-volatile memory, such as One-Time Programmable (OTP) memory.

\item[--] The private key is securely stored within a secure element, such as a TPM or a HSM, integrated within the SoC, ensuring protection against unauthorized access.

\end{enumerate}

The device root key is used for various security functions, including secure boot, device authentication, and secure communication. It also serves as the basis for deriving additional keys, such as session keys and content encryption keys, which are used to secure communication and data. The secure element storing the key is designed to prevent extraction or tampering and allows access only through secure channels.\\

In this setup, TPM technology or secure storage hardware provides a hardware-based layer of security. For instance, a TPM chip includes a dedicated cryptographic processor that safeguards cryptographic keys and digital certificates, offering strong protection against unauthorized access and key compromise.

\subsection{Generation of Private-Public Keys}

In 2019, Keyfactor conducted a study analyzing 75 million digital certificates using RSA keys, finding that 1 in every 172 certificates was vulnerable to attacks capable of exposing the private key~\cite{FactorRSA}. This vulnerability was primarily due to weak random number generation. A significant portion of these compromised certificates was identified in IoT and embedded devices, such as firewalls, routers, and switches. Insufficient randomness in the prime numbers used to generate RSA public keys can result in two distinct keys sharing a common factor, which enables attackers to easily derive the remaining factors and thereby compromise the keys.\\

Public-private key pairs can serve two main purposes: generating and verifying digital signatures, or establishing secure keys for transporting symmetric keys. Each key pair is uniquely associated with an entity, referred to as the key-pair owner. Key pairs can be generated through one of the following methods:

\begin{itemize}
     \item[--] The key-pair owner generates the keys independently in a secure environment, ensuring the secrecy of the private key.

     \item[--] Alternatively, a third party, such as an identity provider, may generate the key pair alongside the owner. When a key pair is intended for use by an individual, it should be protected by a \textit{password} known only to the owner, minimizing the risk of unauthorized use through copying.

\end{itemize}

\subsection{Appropriate certificate issuance process}

An organized and secure certificate issuance process is essential for establishing a trusted TLS environment. This section outlines the recommended steps to ensure that the certificate generated for a server is authenticated, verified, and securely deployed. Each step, from the initial generation of cryptographic key pairs to the installation of the certificate on the server, is designed to maintain the integrity of the server's identity and the security of its communications.

\begin{enumerate}
    
    \item[--] The system administrator (\emph{SysAdmin}) initiates the cryptographic key pair generation process on the TLS server using designated server utilities, thereby creating a public key and a private key by choosing the correct algorithm according to the provided guideline. 

    \item[--] The \emph{SysAdmin} inputs the required domain name into the utilities. This action generates a CSR that encapsulates the server's address and its public key. Subsequently, the SA extracts the CSR from the server and store as a file.

    \item[--] The CSR is then submitted to the legal Registration Authority (RA), who is responsible to review and approve the certificate request.

    \item[--] Upon receiving the CSR, the RA conducts a thorough evaluation, verifying the request's legitimacy and the requester's authorization. Following successful validation, the RA conveys an approval to the Certificate Authority (CA).

    \item[--] The CA, in response to the approval, proceeds to issue the certificate.

    \item[--] The SA is notified by the CA of the certificate's availability, either through an email containing the certificate or via a link for its download. The administrator then acquires the server certificate. Additionally, the SA also obtains the CA certificate chain from the CA.

    \item[--] The acquired server certificate and the certificate chain is then installed on the newly configured server by the SA.

\end{enumerate}    

\subsubsection*{Remarks on Certificate}
\begin{itemize}
    \item[--] As per the NIST guideline~\cite{NIST-SP-800-52-r2}, after 1 January, 2024, only TLS 1.3  shall be used to ensure highest security in transport layer.
    \item[--] Certificates should be signed with same signature algorithms.
    \item[--] The validity period for end-entity certificates should not exceed 398 days (approximately 13 months) to enhance security.
\end{itemize}

\subsection{Compact digital certificates for IoT}

Implementing Public Key Infrastructure (PKI) in Internet of Things (IoT) environments presents challenges due to the size and encoding of standard X.509 public key certificates, which are often too large for constrained devices and networks~~\cite{RFC7228}. To address this, adopting more compact certificate formats, such as the CBOR, offers a solution. CBOR encoding can significantly reduce certificate size, leading to performance benefits such as reduced communication overhead, lower power consumption, decreased latency, and optimized storage—features essential for resource-constrained IoT systems~\cite{RFC8949}. Maintaining compatibility with X.509 certificates during a transitional phase may also be beneficial, given the standard’s widespread use in PKI. By employing efficient encoding formats, IoT implementations can achieve enhanced security and reliability without exceeding device and network limitations.

\subsection{Integrity and Authentication Protection}

Cipher-based Message Authentication Code (CMAC) and Hash-based Message Authentication Code (HMAC) are widely used to provide integrity and authenticity to messages. While both serve similar purposes, they differ in underlying construction, key requirements, security properties, and performance. CMAC can be more efficient in scenarios where block cipher acceleration (such as AES) is available, whereas HMAC offers flexibility and is widely supported across platforms, often providing a higher level of security depending on the application requirements.\\

CMAC relies on a symmetric encryption algorithm, such as AES, to generate a fixed-length tag (typically 128 bits with AES) that is appended to the message. This tag is computed using a shared secret key with the block cipher in a specific mode. The recipient can then verify the integrity of the message by recalculating the tag with the same key. One advantage of CMAC is that it uses the same key as the underlying cipher, simplifying key management.\\

In contrast, HMAC is based on cryptographic hash functions, such as SHA-2 or SHA-3. A separate secret key is used with the hash function to compute a tag whose length depends on the hash function used (e.g., 256 bits for SHA-256). HMAC’s construction ensures that even if the hash function has minor weaknesses, HMAC remains secure, making it a highly reliable choice for a range of applications.

\subsection*{Network Connection Establishment Procedure}

Transport Layer Security (TLS) is the standard cryptographic protocol for authenticating and encrypting communications between clients and servers. TLS relies on digital certificates, which contain verifiable information about the certificate holder’s identity along with an associated private key, essential for establishing secure communication channels.\\

For effective TLS implementation, a server must possess both an authenticated certificate and its corresponding private key. Together, these components enable the server to verify its identity to clients and are crucial for generating symmetric encryption keys. These keys allow the secure encryption and decryption of transmitted data, ensuring confidentiality and integrity in client-server communications. The overall security of the TLS framework relies on two primary factors: secure implementation and configuration of TLS servers, and effective management of TLS certificates. Managing TLS certificates is a complex task that requires continuous monitoring, timely updates, and renewals to prevent security vulnerabilities. Inadequate maintenance practices can lead to serious risks, such as service disruptions, increased susceptibility to cyberattacks, and compromised data integrity.\\

When initiating a secure connection, the client and server engage in a negotiation process to select an appropriate cipher suite. The client initiates this by sending a handshake message containing a list of supported cipher suites. The server then selects one based on its own preference order and responds with a handshake message specifying the chosen suite. It’s important to note that the server may not always select the strongest cipher suite proposed by the client, as it may prioritize differently. Consequently, there’s no guarantee that the most secure cipher suite will be chosen. If no mutually acceptable cipher suite is found, the connection will terminate.\\

To comply with security guidelines, the server must meet specific requirements across several categories, including TLS protocol versions, server keys and certificates, cryptographic functionality, TLS extension support, client verification, session continuation, compression methods, and operational standards.

\begin{itemize}
    \item[--] All device categories (i.e., Standard and Constrained Devices) must use ephemeral keys to provide perfect forward secrecy, without exception. 
    \item[--] The server must perform revocation checking of client certificates whenever client certificate authentication is used. 
    \item[--] Constrained devices that do not support TLS 1.3 should be deployed in configurations that maintain security, even if they use alternative TLS 1.2 versions
    \item[--] TLS 1.3 should not be used with the 0-RTT option unless absolutely necessary, as 0-RTT introduces potential security risks, such as replay attacks. 
\end{itemize}

The following cipher suites shall be used exclusively with elliptic curve-based server certificates. In TLS 1.3 naming conventions, for instance, \texttt{TLS\_AES\_256\_GCM\_SHA384} indicates that messages are encrypted and authenticated using AES-256 in Galois/Counter Mode (GCM), with SHA-384 applied in the HMAC-based Key Derivation Function (HKDF) process.

% The above list is ordered, meaning that the lowest number has the highest preference.

\begin{table}[h!]
    \centering
    \small
    \begin{tabular}{@{}p{0.48\textwidth} p{0.48\textwidth}@{}}
        \hline
        \multicolumn{2}{c}{\textbf{TLS Cipher Suites}} \\
        \textit{Standard Device} & \textit{Constrained Device} \\
        \hline
        \texttt{TLS\_AES\_128\_GCM\_SHA256} & \texttt{TLS\_AES\_128\_GCM\_SHA256} \\
        \texttt{TLS\_AES\_256\_GCM\_SHA384} & \\
        \texttt{TLS\_CHACHA20\_POLY1305\_SHA256} & \\
        \hline
    \end{tabular}
    \caption{Recommended TLS Cipher Suites for Standard Devices and Constrained Devices}
    \label{tab:tls_cipher_recom}
\end{table}

\section{Recommended cipher suite based on operation mode}
\label{sec:2:ciphersuitematrix}

To remain adaptable within the rapidly evolving cryptographic landscape, this architecture includes two operational modes. The first, \textbf{Current Mode}, employs well-established protocols and standards that satisfy our outlined security requirements, providing both robustness and wide deployability. The second, \textbf{Future Mode}, is designed to accommodate emerging algorithms and protocols that have undergone rigorous multi-year evaluations through initiatives such as the NIST Post-Quantum Cryptography Standardization Project. While these algorithms are still being finalized by NIST and other regulatory authorities, adhering to their specifications as they become available will ensure long-term security and compliance.

\subsection{Current Mode}

This mode incorporates cryptographic primitives selected for their broad support across platforms and libraries, as well as extensive analysis and standardization by NIST and other authoritative bodies. These algorithms provide robust security against known attacks while remaining efficient in terms of computation and communication.

\begin{table}[h!]
    \centering
    \begin{tabular}{@{}p{0.5\textwidth} p{0.48\textwidth}@{}}
        \hline
        \multicolumn{2}{c}{\textbf{Current Mode}} \\ 
        \textbf{\textit{Type}} & \textbf{\textit{Specification}} \\
        \hline
        \textbf{Symmetric-key encryption:} & AES-128, AES-256bit GCM~\cite{NIST-GMAC} \\
        \textbf{Public-key algorithms:} & Ed25519 (128 bits), Ed448 (224 bits)~\cite{NIST-ECC-recom} \\
        \textbf{Digital signatures:} &  EdDSA~\cite{NIST-ECC-recom},~\cite{nist-fips-186-5} \\
        \textbf{Integrity \& Authentication:} & CMAC~\cite{NIST-CMAC}, HMAC \\
        \textbf{Key derivation functions:} & As per NIST SP 800-56C~\cite{NIST-800-56C}\\
        \textbf{Random number generation:} & NIST SP 800-90A-C~\cite{nist80090a} \\
        \hline
    \end{tabular}
    \caption{Recommended Cryptographic Algorithms and Standards for \emph{Current Mode}}
    \label{table:crypto-algorithms-cm}
\end{table}

\subsubsection{Recommendations for \emph{Current Mode}}

Public-private key pairs should be generated using elliptic curve cryptography, specifically with Edwards-curve-25519 (Ed25519) and Edwards-curve-448 (Ed448)~\cite{rfc8032}. Ed25519 provides a security level equivalent to 128 bits, while Ed448 offers a security level of approximately 224 bits.\\

The EdDSA (Edwards-curve Digital Signature Algorithm) has multiple advantages, including high performance across diverse platforms and resilience to side-channel attacks. Its compact public key (32 bytes) and signature size (64 bytes) make it particularly suitable for resource-constrained devices. Edwards curves, like Ed25519 and Ed448, are generally preferred over traditional elliptic curves, such as the NIST P-256 and its variants, due to their more robust security properties and simplicity in implementation.\\

A significant advantage of EdDSA is its collision resilience, which means that hash-function collisions do not compromise system security. Additionally, the formulas for Edwards curves are complete, eliminating the need for EdDSA to perform computationally expensive point validation on untrusted public values, thus enhancing both efficiency and security.

\subsection{Future Mode }

The cryptographic primitives for \textbf{Future Mode} are selected based on their performance and security as demonstrated in NIST’s Post-Quantum Cryptography Standardization Project. These include algorithms that are resistant to quantum computing attacks and are expected to serve as foundational standards in the coming years. Given the relatively recent development of these algorithms, they may undergo further changes, so continuous monitoring of updates from NIST and similar bodies is recommended.\\

While NIST indicates that SHA-2 family hash functions, such as SHA-256, SHA-384, and SHA-512, are likely secure beyond 2030, we recommend adopting the SHA-3 family functions for additional resilience against future threats.

%  At present, NIST has chosen one KEM}  for standardization. Four other KEM} are still under evaluation and NIST expects to standardize at least one of them at the end of the fourth round. In this mode, we use the following cryptographic primitives based on the outcome of the Post Quantum Cryptopgraphy competition of NIST~\cite{nist_pqc_project}.  

 \begin{table}[h!]
    \centering
    \begin{tabular}{@{}p{0.55\textwidth} p{0.43\textwidth}@{}}
        \hline
        \multicolumn{2}{c}{\textbf{Future Mode}} \\ 
        \textbf{\textit{Type}} & \textbf{\textit{Specification}} \\
        \hline
        \textbf{Symmetric-key encryption} & AES-256bit GCM  \\
        \textbf{Key Encapsulation Mechanism} & ML-KEM~\cite{nist_fips_203} \\
        \textbf{Digital signatures} & ML-DSA~\cite{nist_fips_204},     SLH-DSA~\cite{nist_fips_205} \\
        \textbf{Hash functions} & SHA-3(384,512)~\cite{nist_fips_202}\\
        \textbf{Random number generation} & NIST SP 800-90A-C~\cite{nist80090a} \\
        \hline
    \end{tabular}
    \caption{Recommended Cryptographic Algorithms and Standards for \emph{Future Mode}}
    \label{tab:primitives_fm}
\end{table}

\section{Conclusion}
In this paper, we laid out a comprehensive framework for securing network-connected devices, targeting critical areas like hardware, firmware, communication, and application security. Starting with secure hardware designs, we highlighted the importance of entropy-based random number generation, secure key management, and root-of-trust implementations. At the firmware layer, techniques such as cryptographic signatures and secure boot processes ensure only validated software runs on devices. For communication, we underlined the necessity of TLS 1.3 and recommended specific cipher suites like AES-GCM and ChaCha20-Poly1305 to meet the needs of both standard and constrained devices. By advocating for compact digital certificates, like CBOR, we addressed the challenges posed by resource-limited IoT systems. Finally, we explored future-proof solutions such as post-quantum cryptographic algorithms to address long-term security concerns in a post-quantum world. Together, these recommendations equip device manufacturers and network architects with the tools needed to build secure, resilient systems that adapt to evolving threats. Moving forward, research efforts will focus on integrating real-time monitoring, automated updates, and intelligent threat detection to strengthen the overall security posture of connected devices.

\bibliography{sec-arch}% common bib file

\end{document}